\documentclass[fleqn]{article}
\begin{document}

\renewcommand{\theequation}{1.\arabic{equation}}
\setcounter{equation}{0}

\title{
{ Asymptotic Pad\'e-Approximant Methods and QCD Current Correlation 
Functions}
}

\author{
{F. Chishtie\thanks{email: fachisht@julian.uwo.ca}, V. Elias\thanks{email: zohar@apmaths.uwo.ca}}\\
{ Department of Applied Mathematics,}
{ The University of Western Ontario,} \\
{ London, Ontario N6A 5B7, Canada. }
\and
T.G. Steele\thanks{email: steelet@sask.usask.ca} \\
{ Department of Physics and Engineering Physics and}\\
{ Saskatchewan Accelerator Laboratory}\\
{ University of Saskatchewan}\\
{ Saskatoon, Saskatchewan S7N 5C6, Canada.}
}
\maketitle

\begin{abstract}
Asymptotic Pad\'e-approximant methods are utilized to estimate the leading-order
unknown ({\it i.e.,} not-yet-calculated) contributions to the perturbative expansions of
two-current QCD correlation functions obtained from scalar-channel fermion and gluon 
currents, as well as from vector-channel
fermion currents.  Such contributions to the imaginary part of each correlator are
polynomials of logarithms whose coefficients (other than the constant term within the 
polynomial) may be extracted from prior-order contributions by use of the 
renormalization-group
(RG) equation appropriate for each correlator.  We find surprisingly good agreement 
between asymptotic Pad\'e-approximant
predictions and RG-determinations of such coefficients for each correlation function 
considered, although such agreement is seen to diminish with increasing $n_f$.  The 
RG-determined coefficients
we obtain are then utilized in conjunction with asymptotic Pad\'e-approximant
methods to predict the RG-inaccessible constant terms of the leading-order
unknown contributions for all three correlators.  The vector channel predictions
lead to estimates for the $O(\alpha_s^4)$ contribution to $R(s) \equiv
[\sigma(e^+ e^- \rightarrow {\rm hadrons}) / \sigma(e^+ e^- \rightarrow \mu^+ \mu^-)]$ 
for three, four, and five flavours.
\end{abstract}

\newpage

\section*{1. Introduction}

     In phenomenological applications of perturbative QCD, one often 
encounters series of the
form $S = 1 + R_1 x + R_2 x^2 + R_3 x^3 + ...$, where $x$ is the expansion 
parameter
$\alpha_s/\pi$, and where only the
first two or three coefficients $R_i$ are known. Since calculation of
the higher-loop diagrams for
subsequent
values of $R_i$ becomes progressively more
difficult, and since the expansion parameter $x$ is often large (as in QCD 
sum rule applications),
it is important to have a reliable method of estimating subsequent 
radiative corrections
to the series $S$ -- that is, of estimating at least the first not-yet-calculated 
coefficient $R_i$.

     In the present paper, we employ the asymptotic Pad\'e-approximant 
approach (already
developed to determine successfully renormalization-group functions within 
QCD [1,2,3] and
scalar
field theories [1,4]) in order to estimate the leading uncalculated 
contributions to the imaginary (absorptive)
parts
of scalar- and vector-current correlation functions within QCD. As all but 
nonlogarithmic terms
within these contributions can be extracted by renormalization group 
methods [3], the asymptotic
Pad\'e-approximant estimates for coefficients of powers of logarithms can 
be tested against
renormalization-group predictions for these coefficients. We find an 
astonishing degree of accuracy
in
such predictions, particularly for the coefficients of the highest and 
next-to-highest powers of
logarithms within the leading uncalculated contributions to correlators.

     In Section 2, we consider the imaginary part of the scalar 
fermion-current correlation
function, which has been calculated [5] to three subleading orders in 
$\alpha_s / \pi$.  The fourth
order term may be expressed as a degree four polynomial in $ln(s/s_0)$, 
where $s_0$ is understood to be the
continuum-threshold parameter of QCD sum-rule integrals in the scalar 
channel. We find good
agreement
between asymptotic Pad\'e-approximant predictions and renormalization-group 
determinations
for the
four coefficients of $ln^k(s/s_0)$ $[k = \{1,2,3,4\}]$ within the 
$[\alpha_s/\pi]^4$ contribution to
the correlator, although
this agreement diminishes as flavour-number $n_f$ increases from 3 to 6.

     In Section 3, we consider the imaginary part of the scalar {\it 
gluon-current} correlation
function,
which has been calculated [6] to two subleading orders in $\alpha_s/\pi$. 
The leading
uncalculated term is a
degree-three polynomial in $ln(s/s_0)$. Asymptotic Pad\'e-approximant 
predictions for all three
coefficients of $ln^k(s/s_0)$ $[k = \{1,2,3\}]$ are found for $n_f\le 4$ to be in 
excellent agreement ($\leq$ 10\% relative error) with values determined from the 
renormalization group, and within 22\% relative error for $n_f=\{5,6\}$.

     In Section 4, the absorptive portion of the vector fermion-current 
correlator is analyzed.
Once again, the leading unknown contribution to this correlation function, 
which has been
calculated
to three subleading orders in $\alpha_s/\pi$ [7], is estimated by 
asymptotic Pad\'e-approximant
methods and
compared to renormalization-group determination of the coefficients of 
$ln^k(s/s_0)$ $[k =
\{1,2,3\}]$. The
agreement is found to be excellent for $k=3$ and $2$, but not for $k=1$, 
where an approximate
factor-of-two discrepancy is seen to occur. This discrepancy is explored 
further by expressing the
asymptotic
Pad\'e-approximant estimate for the leading unknown contribution to the 
vector correlator as an
asymptotic series in the variable $L \equiv -ln(s/s_0)$:  $R_4 = d_3 L^3 + 
d_2 L^2 + d_1 L + ...$ .
The two leading series coefficients $d_{3,2}$ of this
large-L
expansion replicate {\it exactly} the renormalization-group determination 
of the two leading
polynomial
coefficients $d_{3,2}$ within the $R_4$ contribution to the vector 
correlator,
{\it i.e.}, the coefficients of $-ln^3(s/s_0)$ and $ln^2(s/s_0)$. The 
coefficient $d_1$
of $-ln(s/s_0)$ obtained from this large-L expansion, corresponding to an 
alternative Pad\'e
estimate
of the $k=1$ coefficient, is found to differ substantially from the correct 
value.  However, these
two
asymptotic Pad\'e-approximant approaches for estimating the $k=1$ 
coefficient, as delineated
above,
are seen to straddle the correct result, suggesting for the vector 
correlator an overall insensitivity of asymptotic
Pad\'e-approximant methods to the $k=1$ coefficient. A least-squares 
determination of the coefficients $d_k$
confirms this insensitivity to $d_1$.

     Finally, in Section 5 we utilize asymptotic Pad\'e-approximant methods, 
in conjunction with
renormalization-group determinations of $ln^k(s/s_0)$ polynomial 
coefficients, in order to
estimate those
constant ({\it i.e.}, $k = 0$) terms which cannot at present be extracted 
by renormalization-group
methods.   Such terms are of phenomenological interest for 
estimating
higher-loop effects in QCD Laplace and finite-energy sum-rules.
Moreover, such predictions can be tested against subsequent 
perturbative calculations.
Of particular interest are predictions obtained for the 
${\cal{O}}(\alpha_s^4)$ term in
$R(s) \equiv \sigma(e^+ e^- \rightarrow  hadrons)/\sigma(e^+e^- \rightarrow 
\mu^+ \mu^-)$.

\renewcommand{\theequation}{2.\arabic{equation}}
\setcounter{equation}{0}

\section*{2.  The Scalar Correlator for Fermion Currents}

We consider first the scalar current correlation function
\begin{equation}
\Pi_s(p^2) = i \int d^4 y \, e^{i p \cdot y} <0|T j_s (y) j_s (0)|0>,
\end{equation}
based upon the scalar fermion current
\begin{equation}
j_s(y) = \bar\psi (y) \psi (y)
\end{equation}
The absorptive portion of this correlator may be expressed in terms of the 
ratio
$w \equiv s/s_0$, where $s \equiv p^2$, and $s_0$ is the continuum 
threshold usual
to QCD sum-rule analysis (i.e., the kinematic threshold above which 
purely-perturbative QCD alone is
adequate to describe the correlation
function):
\begin{eqnarray}
\frac{1}{\pi} Im \Pi_s & = & \frac{3s}{8 \pi^2}\left[ 1 + \frac{\alpha_s
(s_0)}{\pi}R_1(w) + \left( \frac{\alpha_s (s_0)}{\pi} \right)^2 R_2
(w)\right. \nonumber\\
& + & \left. \left( \frac{\alpha_s(s_0)}{\pi} \right)^3 R_3 (w) + \left(
\frac{\alpha_s (s_0)}{\pi}\right)^4 R_4 (w) + ... \right]
\end{eqnarray}
The coefficients $R_1(w)$, $R_2(w)$ and $R_3(w)$ have been determined
for arbitrary flavour number [5].  For example, when $n_f = 3$, these
are given by
\begin{equation}
R_1(w) = 17/3 - 2 \, ln(w)
\end{equation}
\begin{equation}
R_2(w) = 31.8640 - (95/3) ln \, w + (17/4) (ln \, w)^2
\end{equation}
\begin{equation}
R_3(w) = 89.1564 - 297.596 \, ln \, w + (229/2)(ln \, w)^2 - (221/24)(ln \, 
w)^3.
\end{equation}
This information is sufficient in itself to generate an asymptotic
Pad\'e-approximant prediction for $R_4(w)$ [3,4]:
\begin{equation}
R_4(w) = \frac{R_3^2 (w)\left[ R_2^3(w) + R_1(w) R_2(w) R_3(w) - 2R_1^3
(w) R_3 (w) \right]}{R_2(w)\left[2R_2^3 (w) - R_1^3 (w) R_3 (w) - R_1^2
(w) R_2^2 (w)\right]}
\end{equation}
This prediction, in turn, may be utilized to generate (via numerical
integration) the first five (nonsingular) finite-energy-sum-rule moments
[4]
\begin{equation}
N_k \equiv (k+2) \int_0^1 dw \, w^{k+1} R_4(w)
\end{equation}
of the ${\cal{O}} (\alpha_s^4)$ contribution to the scalar correlation
function.  Substituting (2.7) into (2.8), we find that
\begin{equation}
N_{-1} = 7544.9, \, \, N_0 = 2059.4, \, \, N_1 = 1158.4, \, \, N_2 =
833.47, \, \, N_3 = 673.29
\end{equation}
The significance of these moments is that they may be used to estimate
the polynomial coefficients of $ln(w)$ in $R_4(w)$, which is known to be
fourth-order in $ln \; w$:
\begin{equation}
R_4(w) = d_0 - d_1 \, ln \, w + d_2 (ln \, w)^2 - d_3 (ln \, w)^3 + d_4
(ln \, w)^4;
\end{equation}
Substituting (2.10) into (2.8) we see that
\begin{equation}
N_{-1} = d_0 + d_1 + 2d_2 + 6d_3 + 24d_4
\end{equation}
\begin{equation}
N_0 = d_0 + d_1/2 + d_2/2 + 3d_3/4 + 3d_4/2
\end{equation}
\begin{equation}
N_1 = d_0 + d_1/3 + 2d_2/9 + 2d_3/9 + 8d_4/27
\end{equation}
\begin{equation}
N_2 = d_0 + d_1/4 + d_2/8 + 3d_3/32 + 3d_4/32
\end{equation}
\begin{equation}
N_3 = d_0 + d_1/5 + 2d_2/25 + 6d_3/125 + 24d_4/625.
\end{equation}
The numerical values (2.9) already obtained for these moments lead to
five linear equations in the five unknowns $d_0 - d_4$. Their solution
is
\begin{equation}
d_0 = 252.5, \, \, d_1 = 1339, \, \, d_2 = 1695, \, \, d_3 = 345.7, \,
\, d_4 = 20.38
\end{equation}

The exact values for the coefficients $d_1 - d_4$ may be
extracted [4] from the renormalization group equation (RG)
\begin{equation}
0 = \left[ s_0 \frac{\partial}{\partial s_0} + \beta (x)
\frac{\partial}{\partial x} + 2 \gamma_m (x) \right] Im \, \Pi (L (s_0),
\; x);
\end{equation}
\begin{equation}
x \equiv \alpha_s / \pi; \, \, L(s_0) \equiv ln (s_0 / s) = -ln \, w
; \, \left( s_0 \frac{\partial}{\partial s_0} = \frac{\partial}{\partial L}
\right)
\end{equation}
As is evident from (2.3) and (2.4-6), the correlation function is in the following
form
\begin{eqnarray}
Im \Pi (L, x) & = & \frac{3s}{8\pi} \left[ 1 + (a_0 + a_1 L)x \right. 
\nonumber \\
& + & (b_0 + b_1 L + b_2 L^2) x^2 + (c_0 + c_1L + c_2 L^2 + c_3 L^3) x^3
\nonumber \\
& + & \left. (d_0 + d_1 L + d_2 L^2 + d_3 L^3 + d_4 L^4) x^4 + ... \right]
\end{eqnarray}
where $a_i$, $b_i$ and $c_i$ are known, and where the $d_i$ are unknown.
Given the known $\beta$-and$\gamma$-function coefficients
\begin{equation}
\beta(x) = - (\beta_0 x^2 + \beta_1 x^3 + \beta_2 x^4 + ...),
\end{equation}
\begin{equation}
\gamma_m(x) = -x (1 + \gamma_1 x + \gamma_2 x^2 + \gamma_3 x^3),
\end{equation}
one can obtain via (2.17) the following set of equations for the $d_i$:
\begin{equation}
d_1 = 3 \beta_0 c_0 + 2 \beta_1 b_0 + \beta_2 a_0 + 2 \gamma_3 + 2
\gamma_2 a_0 + 2 \gamma_1 b_0 + 2 c_0
\end{equation}
\begin{equation}
2d_2 = 3 \beta_0 c_1 + 2 \beta_1 b_1 + \beta_2 a_1 + 2 \gamma_2 a_1 + 2
\gamma_1 b_1 + 2c_1
\end{equation}
\begin{equation}
3d_3 = 3\beta_0 c_2 + 2\beta_1 b_2 + 2 \gamma_1 b_2 + 2 c_2
\end{equation}
\begin{equation}
4d_4 = 3\beta_0 c_3 + 2c_3
\end{equation}
For $n_f = 3$ the values of $a_i$, $b_i$, $c_i$ are given in eqs. (2.4-6):
\begin{eqnarray}
a_0 = 5.66667, \; a_1 = 2, \; b_0 = 31.8640, \; b_1 = 31.6667, \; b_2 =
4.25 \nonumber \\
c_0 = 89.1564, \; c_1 = 297.596, \; c_2 = 114.5, \; c_3 = 9.20833
\end{eqnarray}
Corresponding values for $\beta_i$ and $\gamma_i$ are [8]
\begin{eqnarray}
\beta_0 = 2.25, \; \beta_1 = 4, \; \beta_2 = 10.0599, \nonumber \\
\gamma_1 = 3.79166, \; \gamma_2 = 12.42018, \; \gamma_3 = 44.2628.
\end{eqnarray}
One then obtains from eqs. (2.22-25) the following
RG determination of the ${\cal{O}}(\alpha_s^4)$ coefficients
$d_1-d_4$:
\begin{equation}
d_1 = 1563.0, \, d_2 = 1583.6, \, d_3 = 356.04, \, d_4 = 20.143.
\end{equation}
The agreement between these numbers and the asymptotic Pad\'e
predictions (2.16) is astonishing close, ranging from a 14\% relative
error in the prediction of $d_1$ to a 1.2\% error in the prediction of
$d_4$.  This close agreement supports the two-parameter asymptotic error formula [3]
\begin{equation}
\delta_{N+2} \equiv \frac{ R_{N+2}^{[N|1]} - R_{N+2}}{R_{N+2}} = -
\frac{A}{N+1+B}
\end{equation}
used to derive the asymptotic Pad\'e-approximant prediction
(2.7) in ref [4].

The above agreement is not peculiar to $n_f = 3$.  Table 3 tabulates
corresponding results for the $n_f = 4,5$ and $6$ scalar-correlation
functions.  It is evident from Table 3 that the accuracy of asymptotic
Pad\'e-approximant predictions for $d_1 - d_4$ is actually better for
$n_f = 4$ and $5$ than it is for $n_f = 3$.  A diminution of accuracy
becomes evident only when $n_f = 6$.

\section*{3.  The Scalar Correlator for Gluon Currents}

\renewcommand{\theequation}{3.\arabic{equation}}
\setcounter{equation}{0}

We now consider the QCD gluonic scalar current correlation function
$\Pi_G =$ $<\left( \frac{\beta(\alpha_s/\pi) G^2}{\alpha_s \beta_0} 
\right)^2>$.
This correlation function is of the same form as (2.1) but with $j_s (y)$
replaced with the RG-invariant gluonic current $j_G(y)$
\begin{equation}
j_s(y) \rightarrow j_G (y) = \frac{\beta(\alpha_s / \pi)}{\alpha_s
\beta_0} G_{\mu\nu}^{a} (y) G^{a,\mu \nu} (y)
\end{equation}
The absorptive portion of this correlation function can be extracted to
order-$\alpha_s^4$ from a previous three-loop calculation [6]
of the correlation function ${<(G^2)^2>}$.  Using our previous notation we
find that
\begin{eqnarray}
Im \, \, \Pi_G(L, x) & = & \frac{ x^2 }{\pi^2 \beta_0^2}[\beta_0 + \beta_1 x +
\beta_2 x^2 + \beta_3 x^3 ]^2 Im <(G^2)^2> \nonumber\\
& = & \frac{2  x^2 s^2}{\pi^3} \left[ 1 + (a_0 + a_1 L)x
\right. \nonumber \\
& + & (b_0 + b_1 L + b_2 L^2)x^2 \nonumber \\
& + & \left. (c_0 + c_1 L + c_2 L^2 + c_3 L^3) x^3 + ... \right]
\end{eqnarray}
where $x \equiv \alpha_s/ \pi$ and $L \equiv ln (s_0/s)$, as before, and
where the known coefficients $a_0, a_1, b_0, b_1$, and $b_2$ are tabulated 
in Table 4.
These coefficients are sufficient in themselves
to determine the aggregate coefficients of $x$ and $x^2$ in the correlator:
\begin{equation}
R_1(w) = a_0 - a_1 \, ln \, w,
\end{equation}
\begin{equation}
R_2 (w) = b_0 - b_1 \, ln \, w + b_2 [ln (w)]^2,
\end{equation}
\begin{equation}
Im \, \Pi_G = \frac{2 x^2 s^2}{\pi^3} \left[ 1 + x R_1 (w) + x^2 R_2 (w) + x^3 R_3
(w) + ...\right].
\end{equation}
Asymptotic Pad\'e-approximant methods may be utilized to predict that [4]
\begin{equation}
R_3 (w) = \frac{2 R_2^3 (w)}{R_1^3 (w) + R_1 (w) R_2 (w)}.
\end{equation}
This function may then be employed in the integrands of moment integrals
\begin{equation}
P_k \equiv (k+1) \int_0^1 dw \, w^k R_3 (w)
\end{equation}
for $k = \{0,1,2,3\}$.  The numerical values of $P_k$ can then be used,
as before, to determine explicitly the coefficients $c_{0,1,2,3}$ that
characterize $R_3(w)$ as a degree-3 polynomial in $ln(w)$:
\begin{equation}
R_3 (w) = c_0 - c_1 \, ln \, w + c_2 (ln \, w)^2 - c_3 (ln \, w)^3,
\end{equation}
\begin{equation}
P_0 = c_0 + c_1 + 2c_2 + 6c_3
\end{equation}

\begin{equation}
P_1 = c_0 + c_1/2 + c_2/2 + 3c_3/4
\end{equation}

\begin{equation}
P_2 = c_0 + c_1/3 + 2c_2/9 + 2c_3/9
\end{equation}

\begin{equation}
P_3 = c_0 + c_1/4 + c_2/8 + 3c_3/32
\end{equation}
In Table 4, the values of these integrals $P_k$ are tabulated for $n_f =
\{0,2,3,4,5,6\}$, as extracted via (3.6) from (3.3) and (3.4).  These
integrals, in turn are sufficient to predict the coefficients $\{c_0,
c_1, c_2, c_3 \}$ by explicit solution of the set of linear equations
(3.9 - 12).  These predicted values of $c_{0-3}$ are labelled as
asymptotic Pad\'e-approximant predictions (APAP) and tabulated at the
bottom of Table 4.

As in the previous section, all but one $(c_0)$ of these coefficients
may be extracted via the renormalization-group equation satisfied by the 
gluonic
scalar-current correlation function
\begin{equation}
\left( \frac{\partial}{\partial L} + \beta (x) \frac{\partial}{\partial
x} \right) Im \, \Pi_G (L,x) = 0,
\end{equation}
where, as before, $\partial/ \partial L = s_0 \, \partial/ \partial s_0$,
and where $\beta(x)$ is given by (2.20) with the coefficients listed in
Table 2.  Upon application of (3.13) to the explicit form (3.2) for $Im
\, \Pi_G$, one finds that
\begin{equation}
c_1 = 2 \beta_2 + 3 \beta_1 a_0 + 4 \beta_0 b_0
\end{equation}
\begin{equation}
2c_2 = 3 \beta_1 a_1 + 4 \beta_0 b_1
\end{equation}
\begin{equation}
3 c_3 = 4 \beta_0 b_2.
\end{equation}
We have tabulated values of $c_1, c_2, c_3$ extracted via these
equations as $c_1(RGE)$, $c_2 (RGE)$ and $c_3(RGE)$ towards the bottom
of Table 4.  In comparing asymptotic Pad\' e-approximant predictions
(APAP) for $c_{1-3}$ to their actual values, as determined from RG
(3.13), one finds accuracy within 4\% for $c_1$ and $c_2$, and within
6\% for $c_3$ when $n_f = 0$.
This accuracy diminishes somewhat as $n_f$ increases, although the 
agreement remains striking
even out to six flavours.  Table 4's {\it least accurate}
asymptotic Pad\'e-approximant prediction, that of $c_1$ when $n_f = 6$,
differs from the true (RG) value by only a 22\% relative error.
This agreement may be understood to confirm the applicability of the 
one-parameter simplification of the asymptotic
error formula (2.29)
\begin{equation}
\delta_{N+2} = -\frac{A}{N+1}
\end{equation}
obtained in ref. [1] and utilized in Section 5 of ref. [4] to obtain eq. (3.6).

\section*{4.  The Vector Correlator for Fermion Currents}
\renewcommand{\theequation}{4.\arabic{equation}}
\setcounter{equation}{0}

The vector-current correlation functions may be extracted from the 
Adler-function
(``D-function'') presented in ref [7]:

\begin{eqnarray}
\Pi_v (Q^2) & = & \frac{4}{3} \sum_f Q_f^2 \left\{ ln \left( 
\frac{Q^2}{\mu^2}
\right)(1 + x) \right. \nonumber \\
& + & x^2 \left[ A_0 ln \left( \frac{Q^2}{\mu^2} \right) + A_1
\left( ln \left( \frac{Q^2}{\mu^2} \right) \right)^2 \right] \nonumber \\
& + & x^3 \left[ B_0 ln \left( \frac{Q^2}{\mu^2} \right) + B_1
 \left( ln \left( \frac{Q^2}{\mu^2} \right) \right)^2 +
B_2 \left( ln \left( \frac{Q^2}{\mu^2} \right) \right)^3 \right]\nonumber 
\\
& + & \left. {\cal{O}}(x^4) \right\}
\end{eqnarray}
where $x = \alpha_s/ \pi$, as before, and where [7]
\begin{equation}
A_0 = 1.98571 - 0.115295 n_f
\end{equation}
\begin{equation}
A_1 = -1.375 + 0.0833333 n_f
\end{equation}
\begin{equation}
B_0 = -1.23954 ( \sum_f Q_f )^2 / [ 3 \sum_f Q_f^2 ] + 18.2427 - 4.21585 
n_f +
0.0862069n_f^2
\end{equation}
\begin{equation}
B_1 = -8.64820 + 1.04385 n_f - 0.0192159 n_f^2
\end{equation}
\begin{equation}
B_2 = 2.52083 - 0.305556 n_f + 0.00925926 n_f^2
\end{equation}
If we identify $\mu^2$ with the continuum threshold $s_0$, we find that the 
absorptive portion
of this correlator is given by
\begin{equation}
\frac{1}{\pi} Im \Pi_v = -\frac{4}{3} \sum_f Q_f^2 \left[ 1 + x R_1 (w)
+ x^2 R_2(w) + x^3 R_3 (w) + x^4 R_4 (w) ... \right]
\end{equation}
where $w \equiv s/s_0$ and where
\begin{equation}
R_1(w) = 1
\end{equation}
\begin{equation}
R_2(w) = A_0 - (-2A_1) ln \, w
\end{equation}
\begin{equation}
R_3 (w) = (B_0 - \pi^2 B_2) - (-2B_1) ln \, w + 3 B_2(ln \, w)^2.
\end{equation}
With this information, it is possible to repeat the calculation of
Section 2 in order to estimate the ${\cal{O}}(x^4)$ coefficients within
$Im \, \Pi_v$.  One can utilize (4.8), (4.9) and (4.10) within (2.7) to
obtain an asymptotic Pad\'e-approximant prediction of the function
$R_4(w)$.  Using this function, one can explicitly evaluate the moment
integrals $N_{-1}$, $N_0$, $N_1$, and $N_2$, as defined by eq. (2.8).
These four moments are sufficient to provide an estimate of the
polynomial coefficients $d_i$ within $R_4(w)$, which must be degree-3 in
$ln \, w$:
\begin{equation}
R_4(w) = d_0 - d_1 ln \, w + d_2 (ln \, w)^2 - d_3 (ln \, w)^3.
\end{equation}
The four linear equations relating moment integrals $N_k$ to
coefficients $d_i$ are given by (2.11-14), with the constant $d_4$ taken
to be zero [the corresponding $x^4$ term for the scalar correlator was
degree-4 in $ln \, w$].  These four equations can then be solved to
obtain asymptotic Pad\'e-approximant (APAP) estimates of $d_0$, $d_1$,
$d_2$, and $d_3$.

We have tabulated these estimates for $n_f = \{2,3,4,5 \}$ in Table 5.
These estimated values for $d_{1-3}$ can be tested against
renormalization group determinations of these coefficients.
The vector-current correlation function (4.7) can be parametrized similar 
to $\Pi_G
(L, x)$ in (3.2):

\begin{eqnarray}
\frac{1}{\pi} Im \Pi_v (L, x) & = & -\frac{4}{3} \sum_f Q_f^2 \left[ 1 + x 
+
(b_0 + b_1 L) x^2 + (c_0 + c_1 L + c_2 L^2) x^3 \right. \nonumber \\
& + & \left. (d_0 + d_1 L + d_2 L^2 + d_3 L^3) x^4 ...\right]
\end{eqnarray}
where $L \equiv ln (s_0 / s) [=-ln \, w]$, and where the constants in
(4.12) are related to those in (4.1) by
\begin{equation}
b_0 = A_0, \, \, b_1 = -2A_1, \, \, c_0 = B_0 - \pi^2 B_2, \, \, c_1 = -
2B_1, \, \, c_2 = 3B_2.
\end{equation}
The correlator (4.12) has the same RG-invariance as (3.2), and is
therefore a solution of (3.13).  This equation predicts
the following values for $d_1$, $d_2$, and $d_3$:
\begin{equation}
d_1 = 3 \beta_0 c_0 + 2\beta_1 b_0 +\beta_2
\end{equation}
\begin{equation}
d_2 = (3 \beta_0 c_1 + 2 \beta_1 b_1) / 2
\end{equation}
\begin{equation}
d_3 = \beta_0 c_2.
\end{equation}
On can further show via (3.13) [or explicitly from substituting (4.2-5)
into (4.13)] that
\begin{equation}
b_1 = \beta_0, \, \, c_1 = 2 \beta_0 b_0 + \beta_1, \, \, c_2 = \beta_0^2
\end{equation}
in which case
\begin{equation}
d_2 = 3 \beta_0^2 b_0 + \frac{5}{2} \beta_0 \beta_1, \, \, d_3 = \beta_0^3.
\end{equation}

Table 5 displays RG determinations of $d_1$, $d_2$, $d_3$ immediately 
below
their APAP estimates.  Striking agreement is evident from the Table
between APAP and RG estimates of $d_3$ and $d_2$ for $n_f = \{2,3 \}$,
an agreement which deteriorates as $n_f$ increases.  However, the APAP
estimates of $d_1$ seem to be a factor of two too large across the entire 
set of
flavours considered.

This discrepancy, however, is more indicative of insensitivity of APAP
methods to the value of $d_1$ and $d_0$ for the vector correlator case,
rather than of any overall failure of APAP methodology, as epitomized by
eq. (2.7).  Indeed, if we use eq. (2.7) directly (as opposed to the
moment integrals $N_k$) to predict $d_3$, $d_2$, $d_1$ and $d_0$, we not
only obtain much different values for $d_1$, but also results {\it
identical} to the RG determinations of $d_2$ and $d_3$.  To see this,
consider the large-L asymptotic expansion of (2.7), with $R_{1-4}$ as
parametrized in (4.12) [$R_1 = 1, \, R_2 = b_0 + b_1 L, \, R_3 = c_0 +
c_1 L + c_2 L^2$]:
\begin{eqnarray}
R_4 & = & \left[ \frac{c_2^2 (b_1^2 + c_2)}{2 b_1^3} \right] L^3 +
\left[ \frac{c_2^4}{4 b_1^6} -\frac{[c_2^3(1+3b_0)]}{2 b_1^4} + \frac{3
c_1 c_2^2}{2 b_1^3} \right. \nonumber \\
& + & \left. \frac{c_2^2(1- 2b_0)}{4 b_1^2} + \frac{c_1 c_2}{b_1} \right] 
L^2 + {\cal{O}}(L)
\nonumber \\
& \equiv & d_3^{APAP'} L^3 + d_2^{APAP'} L^2 + d_1^{APAP'} L + ...
\end{eqnarray}
Using (4.17) within the first two terms listed in (4.19), we find that
\begin{equation}
d_3^{APAP'} = \beta_0^3,
\end{equation}
\begin{equation}
d_2^{APAP'} = 3 \beta_0^2 b_0 + 5 \beta_0 \beta_1 / 2.
\end{equation}

These APAP$'$ predictions, based entirely upon (2.7), coincide {\it
exactly} with the RG determinations (4.18) of $d_3$ and $d_2$.  In Table
6 we tabulate the corresponding predictions $d_1^{APAP'}$, as obtained
via the large-L expansion (4.19), for the coefficient $d_1$.  Also
tabulated are the prior (APAP) predictions of $d_1$ obtained via use of
(2.7) within the integrands of moment integrals $N_k$, along with
renormalization-group (RG) determinations of $d_1$.  As is evident from the 
Table,
the true (RG) value of $d_1$ is {\it straddled} by the two Pad\'e 
predictions,
demonstrating the insensitivity of Pad\'e methodology to $d_1$, the 
``sub-subleading'' ${\cal{O}}(x^4)$
coefficient in the vector channel.

This insensitivity of the $d_1$ prediction is evident from a 
least-squares approach which finds values for $d_k$ which minimize 
the
quantity
\begin{equation}
\chi^2\left(d_0,d_1,d_2,d_3\right)=\int\limits_0^1\left[R_4(w)-\left(d_0  
-d_1\log(w)+d_2\log^2(w)
-d_3\log^3(w)\right)\right]^2 dw
\end{equation}
where $R_4(w)$ is obtained by substituting (4.8-10) into (2.7).
For $n_f=3$, the $\chi^2$ integral (numerically evaluated) becomes the following 
quadratic form:
\begin{eqnarray}
\chi^2\left(d_0,d_1,d_2,d_3\right)&=&227586.642+720d_3^2+12d_3d_0+24d_2^  
2+48d_1d_3+2d_1^2
+240d_2d_3+d_0^2+12d_1d_2+2d_1d_0
\nonumber\\
& &
+4d_2d_0-237.6405798 d_0-908.3757480 d_1-4402.296530d_1-25479.03758d_3
\end{eqnarray}
Minimization of the $\chi^2$ is equivalent to minimization of the matrix 
quadratic form
\begin{equation}
\left( Ax-b\right)^2
\end{equation}
where
\begin{equation}
x=\left(
\begin{array}{c}
d_0 \\ d_1 \\ d_2 \\ d_3
\end{array}
\right)
\quad ,\quad A=\left(
\begin{array}{cccc}
2 & 2 & 4 & 12 \\
2 & 4 & 12 & 48 \\
4 & 12 & 48 & 240\\
12 & 48 & 240  & 1440
\end{array}
\right)
\quad, \quad
b=\left(
\begin {array}{c}  237.6405798\\ 908.3757480
\\ 4402.296530\\ 25479.03758
\end {array}
\right)
\end{equation}
The singular value decomposition of  $A$ is
\begin{eqnarray}
& &A=U\Sigma V^T\quad , \quad
\Sigma\equiv \sigma_i\delta_{ij}
\\
& & \sigma_0= 1481.970038\quad 
,\quad \sigma_1= 10.52846941 
\quad ,\quad
\sigma_2= 1.395692455\quad ,\quad \sigma_3= 0.1058005879\\
& &U=V=
\left(
\begin {array}{cccc} - 0.008483667052&- 0.2757146361&
 0.8783851152&- 0.3903191722\\- 0.03336418987&-
 0.4565091835& 0.2371489578& 0.8568818869\\-
 0.1652682630&- 0.8311862313&- 0.4115179052&- 0.3353637783
\\- 0.9856476345& 0.1571949543& 0.05341324608&
 0.03058614050\end {array}
\right)
\end{eqnarray}
and the quadratic form becomes
\begin{equation}
\left(Ax-b\right)^2=\left(\Sigma y-b'\right)^2=\sum_{i=0}^3\left(\sigma_i 
y_i-{b'}_i\right)^2
\quad ,\quad b'=U^Tb\quad ,\quad y=V^Tx
\end{equation}
Although  the quadratic form (and $\chi^2$) is minimized by
\begin{equation}
y_i=\frac{b'_i}{\sigma_i}
\Longrightarrow y_0=-17.4587\quad ,\quad y_1=-12.7421\quad ,\quad y_2=-19.0195 \quad ,\quad y_3=-108.2255
\end{equation}   
 the wide range of singular values 
$\sigma_0\gg\sigma_2>\sigma_3$
implies that the $\chi^2$ depends strongly on $y_0$ and $y_1$, but is 
relatively insensitive to $y_2$ and $y_3$.
This broad $\chi^2$ minimum in the $y_2$ and $y_3$  directions compared 
with the sharp minimum
in the $y_0$ and $y_1$ directions implies a comparatively large uncertainty 
in the extraction of $y_2$ and $y_3$ from the Pad\'e approximation. 
Since $y_3$ is both large and uncertain, the relation
 $x=Vy$ implies that $d_0$
 and $d_1$ (and to a lesser extent $d_2$) are
 dominated by the values of $y_2$ and $y_3$ rather than $y_0$ and $y_1$, 
and hence $d_0$, $d_1$ are poorly determined.  
However, the $y_2,~y_3$ 
dependence  of
the RG accessible $d_k$ can be eliminated to find the single linear combination 
independent of $d_0$ that is well determined by the $\chi^2$ minimization:
\begin{equation}
{ d_3}+
 0.01950310416\,{ d_1}+ 0.1410349007\,{ d_2}
= -1.009606933{ y_0}+ 0.0310653405\,{ y_1}=17.23056
\end{equation}
Using the RG values  we find
\begin{equation}
{ d_3(RG)}+
 0.01950310416\,{ d_1(RG)}+ 0.1410349007\,{ d_2(RG)}
=17.17371
\end{equation}
in extremely close agreement (0.33\% relative error) with the Pad\'e prediction (4.31).

The insensitivity of Pad\'e predictions of $d_1$ for the vector correlator  is to be contrasted with 
corresponding predictions for 
the scalar fermionic 
current
correlation function considered in Section 2.  If we apply the large-L
asymptotic expansion of (2.7) to this scalar correlator, with $R_{1-3}$
as given in (2.4-6), we can predict APAP$'$ values for $d_4$, $d_3$,
$d_2$ and $d_1$.  These values are tabulated in Table 7 alongside the
APAP estimates already listed in Table 3. As is evident from the Table,
{\it both} Pad\'e methods predict quite
similar values $d_4$, $d_3$ {\it and} the sub-subleading coefficient $d_2$, 
with the asymptotic large-L
expansion (APAP$'$) approach showing even greater accuracy than the
moment-integral (APAP) approach delineated in Section 2.  For the scalar
correlator, the APAP$'$ approach does not show signs of breaking down until
the third subleading order of (2.7)'s large-L expansion, the coefficient
$d_1$, for which APAP$'$ is considerably less reliable than APAP.
This observation is confirmed  by  a $\chi^2$ minimization analysis of  the scalar correlation function, 
which shows that
the $d_k$ are much less sensitive to 
any poorly determined $y_i$.

Thus we see that different asymptotic Pad\'e estimates do an excellent
job of predicting leading and subleading ${\cal{O}}(x^4)$ coefficients
for both the vector ($d_2, d_3$) and the scalar ($d_3, d_4$) fermionic
current correlators.  However, the approaches diverge drastically in
predicting the sub-subleading term ($d_1$) in the vector correlator,
indicative of the limitations of the Pad\'e method for this case.  By 
contrast,
the sub-subleading term ($d_2$) in the scalar coordinator is quite well
predicted using either method, suggesting greater reliability of APAP
predictions in the scalar channel for subsequent subleading terms.

\newpage

\section*{5. Pad\'e-Estimates of RG-Inaccessible Coefficients}

\renewcommand{\theequation}{5.\arabic{equation}}
\setcounter{equation}{0}

We have seen that the renormalization group equation may be utilized to
determine all but one of the next-order coefficients in current
correlation functions.  Specifically, the coefficient $d_0$ in scalar
(2.19) and vector (4.12) fermion-current correlation functions, as well
as the coefficient $c_0$ in the scalar gluon-current correlation
function (3.2), are not subject to RG constraints.  However, knowledge of
these constants is vital to a number of phenomenological applications, such 
as higher order perturbative contributions to QCD sum-rules, which may be large 
at low $s_0$ [3], or the ${\cal{O}}(\alpha_s^4)$
term in $R \equiv \sigma(e^+ e^- \rightarrow hadrons) /
\sigma (e^+ e^- \rightarrow \mu^+ \mu^-)$.

Up until now, our emphasis has been on demonstrating how
asymptotic Pad\'e-approximant methods can estimate RG-accessible
coefficients within the correlators (2.17), of (3.2), and (4.12).
However, it is evident from
Tables 3,4,5 and 6 that the accuracy of such estimates diminishes as the
subscript index $i$ of coefficients $d_i$ and $c_{i}$ decreases.  The
APAP predictions for $d_0$ in Table 5, for example, are quite suspect
because of the insensitivity of APAP predictions to $d_0$ and $d_1$ in this channel,
as discussed in the previous section.

To obviate these difficulties, we can estimate $d_0$ $(c_0)$ for fermion 
(gluon) correlators
by first
averaging the asymptotic Pad\'e-approximant expression for $R_4 (w)$
[$R_3(w)$] over the interval $0 < w < 1$, and by then removing the known
contributions of RG-accessible coefficients $d_{1-4}$ ($c_{1-3}$) to this 
average.
For the scalar fermion-current correlation function, this procedure
amounts to the use of RG-determined coefficients within (2.11):

\begin{equation}
d_0 = \int_0^1 R_4 (w) dw - d_1 (RGE) - 2d_2 (RGE) - 6d_3 (RGE) -
24d_4 (RGE).
\end{equation}
The integral in (5.1) is just the integral $N_{-1}$ tabulated in Table 3,
where RG-values of $d_{1-4}$ are also listed.  These new estimates
[labelled $d_0$(APAP$'$)] are presented in Table 8 for $n_f$ values of
phenomenological interest, and are compared to the prior estimates of
Table 3 obtained from higher-moment integrals.  The two predictions are
comparable for $n_f = 3,4$, but quite far apart for $n_f = 6$.

Corresponding predictions for the scalar gluon-current
correlation function may be obtained from rearrangement of (3.9):
\begin{equation}
c_0 = \int_0^1 R_3 (w) dw - c_1(RGE) - 2 c_2 (RGE) - 6 c_3 (RGE),
\end{equation}
with $c_{1-3}$(RG) and the integral $(P_0)$ as given in Table 4.
These new predictions for $c_0$, labelled as $c_0$(APAP$'$)
in Table 9, are in good agreement with the previous Table 4 predictions
$c_0$(APAP) for $n_f \leq 4$, reflecting the somewhat better agreement of
Table 4's predictions with RG values.

Finally, $d_0$ estimates for the vector fermionic correlator
\begin{equation}
d_0 = N_{-1} - d_1 (RGE) - 2d_2 (RGE) - 6d_3 (RGE),
\end{equation}
with all constants on the right hand side as tabulated in Table 5,
are displayed in Table 10.  Here the discrepancy with prior APAP estimates 
is quite
large, but within a factor of 2 for the phenomenologically important $n_f = 
5$ case.
We reiterate, however that the Table 5 estimates of $d_0$ are likely
skewed by the factor-of-two disparity between corresponding estimates of
$d_1$ versus RG determinations of $d_1$.  Consequently, we regard the
{\it top line} estimates of Tables 8-10 to be the ones to be tested
against future calculation.

As just one such example, the quantity $R(s)$ is obtained entirely from
the nonlogarithmic coefficients within (4.12):
\begin{eqnarray}
R(s) & \equiv & \sigma(e^+e^- \rightarrow \gamma \rightarrow hadrons) /
\sigma (e^+ e^- \rightarrow \mu^+ \mu^-) \nonumber \\
& = & 3 \sum_f Q_f^2 [1+x+b_0 x^2 + c_0 x^3 + d_0 x^4 ...] .
\end{eqnarray}
The coefficients $b_0$ and $c_0$ are given by (4.13) in terms of constants 
listed in
(4.2-6), and are also tabulated in Table 10.  Thus, for five active flavours we 
predict that
\begin{equation}
R(s) = \frac{11}{3} [1 + x + 1.40924x^2 - 12.8046 x^3 + 31.5 x^4],
\end{equation}
where the well-known first four terms [7,9] have been augmented by the
$n_f = 5$ APAP$'$ estimate for $d_0$ in Table 10.

This prediction for the $x^4$ term in (5.5) differs in both sign and 
magnitude
from an earlier Pad\'e-motivated prediction [10].  The prediction is also
substantially less than the 73.5$x^4$ term one would
obtain applying (2.7) directly to the first three terms in (5.5). Such an 
approach, however, is
based on the constants $b_0$ and $c_0$ only, corresponding to finding 
$R_4(w)$ only
for the specific choice $w=1$ within (2.7).  By contrast, the result (5.5) 
devolves from an estimate incorporating our exact
knowledge of all coefficients $b_i$, $c_i$, and $d_i$ (except $d_0$) to 
find the
average value of $R_4(w)$ over the full range of $w$.

     In Table 10, the exact values for $b_0$ and $c_0$ are tabulated
for $n_f = \{2,3,4\}$ as well.  These values
can be substituted into (5.4) to display $n_f = \{3,4\}$ expressions for 
$R(s)$
which incorporate the APAP$'$ estimate for $d_0$. It is
interesting to note that the predicted magnitude of the $x^4$ coefficients 
in $R(s)$
is quite modest, suggesting that the accuracy of phenomenology based on the 
preceding three subleading orders
of perturbation theory is not compromised by higher-order corrections.

\section*{Acknowledgment}

We are grateful for research support from the Natural Sciences and 
Engineering Research
Council of Canada.

\section*{References}

\begin{enumerate}

\item{} J. Ellis, M. Karliner, and M. A. Samuel, Phys. Lett. B 400, 176 
(1997)
\item{} J. Ellis, I. Jack, D.R.T. Jones, M. Karliner, and M. A. Samuel, 
Phys. Rev. D 57, 2665
(1998)
\item{} V. Elias, T. G. Steele, F. Chishtie, R. Migneron, and K. Sprague, 
Phys. Rev. D 58, 116007(1998)
\item{} F. A. Chishtie, V. Elias, and T. G. Steele, Phys. Lett. B (to 
appear) hep-ph/9809538
\item{} K. G. Chetyrkin, Phys. Lett. B 390, 309 (1997)
\item{} K. G. Chetyrkin, B. A. Kniehl, and M. Steinhauser, Phys. Rev. Lett. 
79, 353 (1997)
\item{} S. G. Gorishny, A. L. Kataev, and S. A. Larin, Phys. Lett. B 259, 
144 (1991)
\item{} T. van Ritbergen, J. A. M. Vermaseren, and S. A. Larin, Phys. Lett. 
B 400, 379 (1997) and
B 405, 323 (1997);  K. G. Chetyrkin, Phys. Lett. B 404, 161 (1997)
\item{} L. R. Surguladze and M. A. Samuel, Phys. Rev. Lett. 66, 560 (1991)
\item{} M. A. Samuel, J. Ellis, and M. Karliner, Phys. Rev. Lett. 74, 4380 
(1995);\\
 A.L. Kataev, V.V. Starshenko, Mod. Phys. Lett. {\bf A10} (1995) 235.

\end{enumerate}

\newpage
\begin{table}
\begin{center}
\begin{tabular}{cccc}
$$ & $n_f=4$ & $n_f = 5$ & $n_f = 6$\\
\hline
$a_0$ & 17/3 & 17/3 & 17/3 \\
$a_1$ & 2 & 2 & 2 \\
$b_0$ & 30.5054 & 29.14671 & 27.7881 \\
$b_1$ & 30.4444 & 29.2222 & 28.0000 \\
$b_2$ & 4.08333 & 3.91667 & 3.75000 \\
$c_0$ & 65.1980 & 41.7576 & 18.8351 \\
$c_1$ & 267.589 & 238.381 & 209.970 \\
$c_2$ & 104.384 & 94.6759 & 85.3750 \\
$c_3$ & 8.39352 & 7.61574 & 6.87500
\end{tabular}
\caption{Coefficients of ${\cal{O}}(\alpha_s)$, ${\cal{O}}(\alpha_s^2)$ and 
${\cal{O}}(\alpha_s^3)$
in terms within the fermionic scalar-current correlation function, as
defined in (2.1).}
------------------------------------------------------------------------  
---------------------------
\end{center}
\end{table}

\begin{table}
\begin{center}
\begin{tabular}{ccccccc}
$$ & $n_f = 0$ & $n_f = 2$ & $n_f = 3$ & $n_f = 4$ & $n_f = 5$ & $n_f =
6$ \\
\hline
$\beta_0$ & 11/4 & 29/12 & 9/4 & 25/12 & 23/12 & 7/4 \\
$\beta_1$ & 51/8 & 115/24 & 4 & 77/24 & 29/12 & 13/8 \\
$\beta_2$ & $\frac{2857}{128}$ & $\frac{48241}{3456}$ &
$\frac{3863}{384}$ & $\frac{21943}{3456}$ & $\frac{9769}{3456}$ & $-
\frac{65}{128}$ \\
$\gamma_1$ & - & 3.93056 & 3.79166 & 3.65277 & 3.51389 & 3.37500 \\
$\gamma_2$ & - & 14.8393 & 12.4202 & 9.94702 & 7.41985 & 4.83866 \\
$\gamma_3$ & - & 61.8794 & 44.2688 & 27.3088 & 11.0401 & -4.50240
\end{tabular}
\caption{Coefficients of QCD $\beta$- and $\gamma$-functions, as defined
in (2.20) and (2.21).}
------------------------------------------------------------------------  
---------------------------
\end{center}
\end{table}

\begin{table}
\begin{center}
\begin{tabular}{cccc}
$$ & $n_f=4$ & $n_f=5$ & $n_f=6$ \\
\hline
$N_{-1}$ & 6217.2 & 5050.4 & 4045.6 \\
$N_0$ & 1616.9 & 1238.0 & 920.15 \\
$N_1$ & 874.58 & 634.83 & 437.31 \\
$N_2$ & 609.96 & 422.70 & 270.86 \\
$N_3$ & 480.56 & 320.06 & 191.81 \\
\\
$d_0$ (APAP) & 147.4 & 64.18 & 9.004 \\
\\
$d_1$ (APAP) & 1029.6 & 744.9 & 442.2 \\
$d_1$ (RGE) & 1159.8 & 791.52 & 457.39 \\
\\
$d_2$ (APAP) & 1398 & 1177 & 1067 \\
$d_2$ (RGE) & 1338.9 & 1114.7 & 910.31 \\
\\
$d_3$ (APAP) & 307.4 & 253.1 & 174.3 \\
$d_3$ (RGE) & 305.7 & 260.06 & 218.82 \\
\\
$d_4$ (APAP) & 16.63 & 15.37 & 17.29 \\
$d_4$ (RGE) & 17.312 & 14.755 & 12.461
\end{tabular}
\caption{Comparison of asymptotic Pad\'e-approximant predictions (APAP)
to renormalization-group equation (RGE) determinations of the
${\cal{O}}(\alpha_s^4)$ coefficients $d_1 - d_4$ within the fermionic
scalar-current correlation function.  The integrals $N_i$, as defined by
(2.8) are obtained numerically from the APAP estimate (2.10) of $R_4$, with
the lower-order coefficients $a_i$, $b_i$, $c_i$ (Table 1) determining
$R_{1,2,3}$.  APAP estimates of $d_0 - d_4$ are obtained from these 
integrals via (2.11-15).
RGE determinations of $d_1 - d_4$ are obtained via (2.22-25), with 
$\beta_i$ and $\gamma_i$
values as given in Table 2.}
------------------------------------------------------------------------  
---------------------------
\end{center}
\end{table}

\begin{table}
\begin{center}
\begin{tabular}{ccccccc}
$$ & $n_f=0$ & $n_f = 2$ & $n_f =  3$ & $n_f = 4$ & $n_f=5$ & $n_f = 6$
\\
\hline
\\
$a_0$ & $\frac{1007}{44}$ & $\frac{6919}{348}$ & $\frac{659}{36}$ &
$\frac{4999}{300}$
& $\frac{4123}{276}$ & $\frac{367}{28}$\\
\\
$a_1$ & $\frac{11}{2}$ & $\frac{29}{6}$ & $\frac{9}{2}$ & $\frac{25}{6}$ &
$\frac{23}{6}$
& $\frac{7}{2}$ \\
\\
$b_0$ & 349.140 & 246.434 & 197.515 & 150.210 & 104.499 & 60.3685 \\
\\
$b_1$ & $\frac{3225}{16}$ & $\frac{7379}{48}$ & $\frac{2105}{16}$ & 
$\frac{1769}{16}$
&
$\frac{4355}{48}$ & $\frac{1153}{16}$\\
\\
$b_2$ & $\frac{363}{16}$ & $\frac{841}{48}$ & $\frac{243}{16}$ & 
$\frac{625}{48}$ &
$\frac{529}{48}$ & $\frac{147}{16}$\\
\\
$P_0$ & 11167.0 & 6769.0 & 4970.7 & 3437.6 &2173.9  & 1191.2\\
$P_1$ & 6966.7 & 4052.1 & 2870.2 & 1875.5  & 1075.7 & 486.57 \\
$P_2$ & 5916.2 & 3382.6 & 2359.5 & 1503.6 & 823.73 & 336.41 \\
$P_3$ & 5449.3 & 3087.2 & 2135.7 &1342.5  & 716.80 & 275.47 \\
\\
$c_0$ (APAP) & 4262 & 2346 & 1580 & 950.4 & 466.0 & 144.5\\
\\
$c_1$ (APAP) & 4148 & 2559 & 1895 & 1311 & 805.8 & 377.5 \\
$c_1$ (RG) & 4323.5 & 2695.9 & 2017.4 & 1424.8 & 915.12 & 485.46 \\
\\
$c_2$ (APAP) & 1145 & 775.4 & 622.8 & 490.2 & 376.3 & 278.6\\
$c_2$ (RG) & 1161.2 & 777.76 & 619.03 & 480.73 & 361.69 & 260.75 \\
\\
$c_3$ (APAP) & 77.92 & 52.21 & 41.71 & 32.63 & 24.92 & 18.65 \\
$c_3$ (RG) & 83.19 & 56.46 & 45.56 & 36.17 & 28.16 & 21.44
\end{tabular}
\caption{The first five rows list known terms of the gluonic scalar-current 
correlation function (3.2).  The integrals $P_{0-3}$, as defined by
(3.7), are obtained numerically and utilized to obtain asymptotic
Pad\'e-aproximant (APAP) estimates of the next-order coefficients $c_{0-
3}$ in the gluonic correlator.  Renormalization-group equation (RG)
determinations of $c_1$, $c_2$, and $c_3$ are also tabulated to illustrate
the accuracy of the APAP method.}
------------------------------------------------------------------------  
---------------------------
\end{center}
\end{table}

\begin{table}
\begin{center}
\begin{tabular}{ccccc}
$$ & $n_f = 2$ & $n_f = 3$ & $n_f = 4$ & $n_f = 5$\\
\hline
$N_{-1}$ & 160.31 & 118.82 & 86.172 & 63.940\\
$N_0$ & 24.561 & 20.699 & 21.182 & 26.106\\
$N_1$ & 11.817 & 13.936 & 19.911 & 29.596\\
$N_2$ & 10.173 & 14.315 & 22.373 & 34.181\\
\\
$d_0$(APAP) & 20.51 & 27.54 & 40.22 & 58.83\\
\\
$d_1$(APAP) & -79.88 & -83.30 & -98.87 & -130.3\\
$d_1$(RGE) & -35.490 & -46.238 & -56.903 & -63.989\\
\\
$d_2$(APAP) & 66.11 & 51.93 & 49.19 & 62.01\\
$d_2$(RGE) & 59.701 & 47.404 & 36.561 & 27.111\\
\\
$d_3$(APAP) & 14.57 & 11.78 & 7.742 & 1.899 \\
$d_3$(RGE) & 14.114 & 11.391 & 9.0422 & 7.041
\end{tabular}
\caption{Comparison of asymptotic Pad\'e-approximant predictions (APAP) to
renormalization-group equation(RGE) determinations of the 
${\cal{O}}(\alpha_s^4)$
coefficients $d_i$ within the fermionic vector-current correlation 
function.
The integrals $N_k$, as defined by (2.8), are obtained numerically from
the APAP estimate (2.7) of $R_4$. APAP estimates of $d_0 - d_3$ are
obtained from these integrals, as described in the text.}
------------------------------------------------------------------------  
---------------------------
\end{center}
\end{table}

\begin{table}
\begin{center}
\begin{tabular}{ccccc}
$$ & $n_f = 2$ & $n_f = 3$ & $n_f = 4 $ & $n_f = 5$ \\
\hline
$d_1$(APAP$'$) & -11.48 & -24.20 & -36.33 & -44.84 \\
$d_1$(RG) & -35.490 & -46.238 & -56.903 & -63.989 \\
$d_1$(APAP) & -79.88 & -83.30 & -98.87 & -130.3 \\
\end{tabular}
\caption{Pad\'e estimates $d_1$(APAP$'$) for
the vector correlator, as obtained via large-L asymptotic expansion of 
(2.7),
are compared to the exact results $d_1$(RG) and to the Pad\'e estimates
$d_1$(APAP) of Table 5 obtained through use of (2.7) in integrands of
moment integrals $N_k$ (2.8).  The two Pad\'e estimates are
seen to straddle the exact result.}
------------------------------------------------------------------------  
---------------------------
\end{center}
\end{table}

\begin{table}
\begin{center}
\begin{tabular}{ccccc}
$$ & $n_f = 3$ & $n_f = 4$ & $n_f = 5$ & $n_f = 6$\\
\hline
$d_4$(APAP$'$) & 20.183 & 17.323 & 14.745 & 12.434 \\
$d_4$(RGE) & 20.143 & 17.312 & 14.756 & 12.461 \\
$d_4$(APAP) & 20.38 & 16.63 & 15.37 & 17.29 \\
\\
$d_3$(APAP$'$) & 355.5 & 305.0 & 259.2 & 217.9 \\
$d_3$(RGE) & 356.03 & 305.73 & 260.06 & 218.82 \\
$d_3$(APAP) & 345.7 & 307.4 & 253.1 & 174.3 \\
\\
$d_2$(APAP$'$) & 1617 & 1380 & 1111 & 916.2 \\
$d_2$(RGE) & 1583.6 & 1338.9 & 1114.7 & 910.31 \\
$d_2$(APAP) & 1695 & 1398 & 1177 & 1067 \\
\\
$d_1$(APAP$'$) & 1450 & 244.7 & 2582 & 405.6 \\
$d_1$(RGE) & 1563.0 & 1159.8 & 791.52 & 457.39 \\
$d_1$(APAP) & 1339 & 1030 & 744.9 & 442.2\\
\end{tabular}
\caption{Pad\'e estimates $d_i$(APAP$'$) of coefficients $d_i$ for the 
fermionic
scalar-current correlator, as obtained via large-L asymptotic expansion of 
(2.7),
are compared to exact (RGE) results and to the APAP estimates of Table 3. 
 The APAP$'$
estimates are seen to predict $d_4$, $d_3$, and $d_2$ with even better 
accuracy than the APAP
estimates, but to suffer substantially diminished accuracy in predicting 
$d_1$.}
------------------------------------------------------------------------  
---------------------------
\end{center}
\end{table}

\begin{table}
\begin{center}
\begin{tabular}{ccccc}
$$ & $n_f = 3$ & $n_f = 4$ & $n_f = 5$ & $n_f = 6$ \\
\hline
$d_0$(APAP$'$) & 195 & 130 & 115 & 156 \\
$d_0$(APAP) & 252 & 147 & 64.2 & 9.00
\end{tabular}
\caption{Asymptotic Pad\'e-approximant predictions of the
$d_0$-coefficient in the scalar fermion-current correlation function.
$d_0$(APAP$'$) is obtained from (5.1) using renormalization-group
determinations of $d_{1-4}$, as given in Table 3.  $d_0$(APAP) are
previous predictions obtained in Table 3 without RG-inputs.}
------------------------------------------------------------------------  
---------------------------
\end{center}
\end{table}

\begin{table}
\begin{center}
\begin{tabular}{ccccccc}
$$ & $n_f = 0$ & $n_f = 2$ & $n_f = 3$ & $n_f = 4$ & $n_f = 5$ & $n_f =
6$\\
\hline
$c_0$(APAP$'$) & 4022 & 2179 & 1442 & 834.3 & 366.4 & 55.6 \\
$c_0$(APAP) & 4262 & 2345 & 1580 & 950.4 & 466.0 & 144.5
\end{tabular}
\caption{Asymptotic Pad\'e-approximant predictions of the
$c_0$-coefficient in the scalar gluon-current correlation function.
$c_0$(APAP$'$) is obtained from (5.2) using renormalization-group
determinations of $c_{1-3}$, as given in Table 4.  $c_0$(APAP) are
previous predictions obtained in Table 4 without RG-inputs.}
------------------------------------------------------------------------  
---------------------------
\end{center}
\end{table}

\begin{table}
\begin{center}
\begin{tabular}{ccccc}
$$ & $n_f = 2$ & $n_f = 3$ & $n_f = 4$ & $n_f = 5$ \\
\hline
$d_0$(APAP$'$) & -8.29 & 1.90 & 15.7 & 31.5 \\
$c_0$ & -9.14051 & -10.2839 & -11.6856 & -12.8046\\
$b_0$ & 1.75512 & 1.63982 & 1.52453 & 1.40924\\
\\
$d_0$(APAP) & 20.5 & 27.5 & 40.2 & 58.8
\end{tabular}
\caption{Asymptotic Pad\'e-approximant predictions of the $d_0$-coefficient 
in the vector fermion-current correlation function.  $d_0$(APAP$'$) is 
obtained from
(5.3) using renormalization-group determinations of $d_{1-3}$, as given in 
Table 5.
$d_0$(APAP) are previous predictions obtained in Table 5 without RG-inputs.
Exact values of $b_0$ and $c_0$ [7,9] are also displayed to facilitate use 
of (5.4) to obtain
$n_f = 3,4$ expressions for $R(s)$.}
------------------------------------------------------------------------  
---------------------------
\end{center}
\end{table}

\end{document}